\documentstyle[aps,epsf,preprint]{revtex}

\def\1{{\bf 1}}
\def\[{\left[}
\def\]{\right]}
\def\be{\begin{eqnarray}}
\def\ee{\end{eqnarray}}
\def\nn{\nonumber}
\def\({\left(}
\def\){\right)}
\def\bk#1{\langle#1\rangle}

\def\labels#1{\hbox{\hspace{.2in}$_{#1}$} \label{#1}}
\def\eq#1{(\ref{#1})}

\def\o{\omega}

\begin{document}
\def\h{{1\over 2}}
\def\s{\sigma}
\def\l{\lambda}
\def\v#1{\vec #1}
\def\.{\cdot}
\def\e{\epsilon}

\title{Implementing Unitarity in Perturbation Theory}
\author{C.S. Lam}
\address{Department of Physics, McGill University\\
 3600 University St., Montreal, Q.C., Canada H3A 2T8\\
Email: Lam@physics.mcgill.ca}
\date{\today}
\maketitle

\begin{abstract}
Unitarity cannot be perserved order by order in ordinary perturbation
theory because the constraint $UU^\dagger=\1$ is nonlinear.
However, the corresponding constraint for $K=\ln U$, being $K=-K^\dagger$,
is linear so it can be maintained in every order in
a perturbative expansion of $K$. 
The perturbative expansion of $K$
may be considered as a non-abelian generalization of the linked-cluster
expansion in probability theory and in
statistical mechanics, and possesses similar advantages
resulting from separating the short-range correlations from long-range
effects.
This point is illustrated in two QCD examples, in which 
delicate cancellations encountered in summing Feynman diagrams
of are avoided when they are calculated via the perturbative
expansion of $K$.
Applications to other problems are briefly discussed.
\end{abstract}
\narrowtext

Probability conservation is not maintained order by order in 
ordinary perturbation
theory. This happens because the unitarity relation
$UU^\dagger=\1$ (for the time-evolution operator $U$) 
is nonlinear, whereas the constraints for the other  
exact conservation laws, such
as energy, momentum, and charge, are linear. 
Order-by-order probability conservation can be restored
if we expand instead $K=\ln U\equiv \sum_{n\ge 1}K_n/n$, for then 
the unitarity constraint becomes the linear constraint $K=-K^\dagger$.
As long as every $K_n$ is kept to be
anti-hermitean, $U=\exp(K)$ will be
unitary no matter where the $K$-expansion is truncated. 
We shall henceforth refer to the perturbation theory in which $K=\ln U$
is expanded as the {\it unitary perturbation theory}. Mathematically,
such an expansion is known as the Magnus expansion \cite{MAGNUS}.

Besides unitarity, there is another reason to deal with the 
perturbative expansion $K_n$ of $K=\ln U$, rather than 
$U=\1+\sum_{n\ge 1}U_n$, because the former is the generalization
of the linked-cluster expansion in probability theory and in
statistical mechanics \cite{HUANG}, 
with the associated advantages in separating
short-range correlations from the long-range effects. 
It is also analogous to the expansion of the effective potential in
quantum field theory.
The added complication here is that we are dealing with non-commutative
operators rather than c-number functions. 
In QCD, short-range correlations refer to correlations in time and in 
color. This point will be explained later with two QCD examples.

Unitary perturbation theory can also be regarded as an extension
of the familiar eikonal approximation. 
For elastic scattering of a high-energy particle from a static
potential, the scattering amplitude is diagonal in energy
and the impact parameter. The scattering amplitude $\bk{U}$  
to all orders can be summed up to
the form $\bk{\exp(K_1)}=\exp(\bk{K_1})$, where 
$\bk{K_1}$ is equal to the first Born approximation
$\bk{U_1}$, and $\bk{\cdots}$
is the matrix element in impact-parameter space \cite{SAK,KF}.
 This eikonal approximation is also valid
for electron-electron scattering when fermion loops
are neglected \cite{EIK}. For high-energy scattering of a quark from a color
potential, amplitudes of all orders can again be summed up into
the form $\bk{\exp(K)}=\exp(\bk{K})$, but now $\bk{K}$ 
is a color matrix, with $\bk{K_n}$
proportional to the $n$th nested commutator of the color matrices $t_a$
\cite{FHL}.
In particular, $\bk{K_n}\not=0$ even for $n>1$. Therefore 
unitary perturbation theory can be thought of as a generalization 
of the non-abelian eikonal approximation.

The relation between $U_n$ and $K_n$ can be obtained
by expanding the exponential $\exp(\sum K_n/n)$ to the $n$th order. 
One obtains
\be
U_n=\sum_{(m)}{1\over k!}{1\over\prod_{i=1}^km_i}K_{m_1}K_{m_2}\cdots K_{m_k},
\labels{unk}\ee
where the sum is taken over all partition $(m)=(m_1m_2\cdots m_k)$
of the integer $n$. If $U$ is the time-evolution operator between times
$T'$ and $T$, ordinary perturbation theory tells us
\be
U_n
&=&(-i)^n\int_{{\cal R}_n}d^nt\ H_I(t_1)\cdots H_I(t_n),\labels{uh}\ee
where  ${\cal R}_n$ is the 
hyper-triangular integration region $\{T\ge t_1\ge\cdots t_n\ge T'\}$.
Together with \eq{unk}, this may be used to derive a dynamical expression
for $K_n$.
In particular,
\be
K_1&=&U_1=-i\int_{T'}^{T}dt\ H_I(t),\nn\\
K_2&=&2U_2-U_1^2=(-i)^2\int_{T'}^Tdt_1\int_{T'}^{t_1}dt_2\ [H_I(t_1),H_I(t_2)].\labels{k12}\ee

The difference 
between ordinary and unitary perturbation theories can be illustrated
by a simple spin-$\h$ example \cite{ALI}, with
interaction $-\h\v\s\.\!\v B(t)$, and an external magnetic field 
 $\v B(t)=(B_\perp\cos\o t,-B_\perp\sin\o t,B_0)$. The longitudinal
field $B_0$ causes a Zeeman splitting of magnitude
$B_0=\hbar\o_0$; the transverse field $B_\perp$ induces a transition
between the lower and the upper states.
Assuming the system to occupy the lower state initially, 
the probability to be in the upper state
is plotted in Fig.~1, as a function of
 the scaled
frequency variable $\Omega=\hbar(\o-\o_0)/B_\perp$ at the scaled time  $\tau=B_\perp t/\hbar=\pi$. The solid, deahed, dash-dot, and dotted
curves are respectively the exact solution, the first Born approximation
$P_1=\bk{U_1}$, the first unitary approximation $UP_1=\bk{\exp(K_1)}$, and the
second unitary approximation $UP_2=\bk{\exp(K_1+K_2/2)}$. 
 The notation
$\bk{\cdots}$ represents the matrix element between
the upper and the lower states.
We see from Fig.~1 
that the probability given by
the Born approximation is larger than 1
for small $|\Omega|$, violating
unitarity, but both unitary approximations stay within
the unitarity bound. As a function of $\tau$
at the resonance frequency $\Omega=0$, the exact solution as well
as the two unitary approximations are periodic, with a period $2\pi$
oscillating between 0 and 1, whereas the Born approximation
grows monotonically like $\tau^2$, making it a worse and worse
 approximation at large $\tau$.

For QCD, the unitary perturbation theory can be used to classify
unitary parton-parton elastic amplitudes and to produce unitary
model for Pomeron amplitudes \cite{FDKL}.

Besides unitarity, it is often
more profitable to compute $U_n$ via $K_n$, 
because $K_n$ contains short-time and color correlations
via the appearance of the
commutator of the $H_I$'s, instead of their product in $U_n$.
I will illustrate
this remark with two second-order QCD
examples. Higher-order situtations will
be discussed later.

The first example concerns high-energy electron-electron and quark-quark 
elastic scattering near the forward direction, 
with Mandelstam variables $s$ and $t$. 
The only difference between the two cases is the presence of
a color matrix $t_a$ at the vertices of QCD diagrams, Fig.~2(a),(b),(c). 
The formulas
for quark-quark amplitudes \cite{CW} are equally applicable to 
electron-electron amplitudes if we replace $t_a$ by 1. 
All amplitudes have a common factor $is/2m^2$ which will be factored out,
leaving the rest to be $\bk{U}$.
Fig.~2(a)
gives the first Born amplitude $\bk{U_1}=(ig^2/\Delta^2)G_1$, where
$\Delta$ is the (transverse) momentum transfer with $t=-\Delta^2$,
and
$G_1=t_a\times t_a$ is the color factor of Fig.~2a. 
 The amplitude
for Fig.~2(b) is $(-ig^4/2\pi)I_2\ln(se^{-\pi i})G_2$, and that 
for Fig.~2(c) is $(ig^4/2\pi)I_2\ln s(G_2+G_1N_c/2)$, where
$I_2=\int d^2k_\perp/[(2\pi)^2k_\perp^2(\Delta-k_\perp)^2]$.
The color factor $G_2=G_1\times G_1$ is that of Fig.~2(b), with $N_c$ being the number
of colors in the theory. This number is to be set equal to 0 for
QED. The second Born amplitude, being the sum of Figs.~2(b) and 
2(c), is $\bk{U_2}=(-ig^4/2\pi)I_2(-\pi i G_2-
\ln s G_1N_c/2)$. Note the cancelation of the $\ln s$ factor in the
term proportional to $G_2$.
The $G_1$ term consists of a color-octet exchange.
It is not negligible compared to the first Born amplitude $\bk{U_1}$
in the leading-log approximation where $g^2\ll 1$ and $g^2\ln s=O(1)$. 
In fact it is the beginning 
of the reggeized gluon contribution. 
The color factor $G_2=G_1\times
G_1$ consists of two color-octet exchanges, which can
be decomposed into a color-octet and a color-singlet.
Its coefficient
in the impact-parameter representation is equal to the coefficient
of $G_1\times G_1$ in $\h\bk{U_1^2}$, a factor of $\ln s$ down  
from the coefficient of the $G_1$
term. Its color-octet contribution is therefore negligible in 
the leading-log approximation, but its color-singlet component is
leading and constitute the beginning of the Pomeron amplitude.
 For QED, the factor $N_c$ in front of
$G_1$ is zero, and $G_2$ is replaced by 1. Hence
$\bk{U_2}=\h\bk{U_1}^2$ in the impact-parameter
representation.

To summarize, a delicate cancelation of $\ln s$ occurs
between Figs.~2(b) and 2(c)
in QED, and in the color-singlet channel of QCD. 
This is reminiscent of the delicate cancellation of the volume
factor in the grand partition function in statistical mechanics
unless the linked-cluster expansion is used \cite{HUANG}.
As a result, the $k$ color-octet exchange
amplitude in QCD is proportional to $g^{2k}(g^2\ln s)^{2-k}\ (k=1,2)$,
and it is this dependence that leads to the reggeon amplitudes in QCD.
We shall now see that if we calculate $\bk{U_2}$ via 
$\bk{K_1^2}$ and $\bk{K_2}$, then
delicate cancelation is not needed,
and the energy dependence needed for the 
reggeon structure becomes immediate.

Since $\bk{K_1}=\bk{U_1}$, it is given simply by Fig.~2(a). 
For $\bk{K_2/2}$, it can be computed from Fig.~2(c), provided
the product of the color matrices $t_at_b$ in the upper quark line
is replaced by their commutator $[t_a,t_b]$ \cite{FHL}, thus
making the color factor purely $G_1$. Otherwise, the result is
identical to that of Fig.~2(c).
In the impact-parameter representation where $K_i$ are 
diagonal, we obtain in this way the same result
as before: $\bk{K_2/2}$ is given by the $G_1$ coefficient of 
Fig.~2(c), or that of
$\bk{U_2}$, and $\bk{K_1}^2/2$ is given by the $G_2=G_1\times G_1$
 coefficient of $\bk{U_2}$. 
The absence of the $\ln s$ factor in $\bk{K_1}^2/2$ is not due to 
cancelation; it is simply a consequence that $\bk{K_1}$ is independent
of $s$. This is the same as in statistical mechanics when linked
cluster expansion is used \cite{HUANG}.
The reggeized gluon contribution is now isolated in $\bk{K_2/2}$
alone, without having to combine two Feynman diagrams to obtain it.
Its octet origin can be traced back to the color commutator in the upper
quark line, or the commutator structure of $K_2$ shown in \eq{k12}.
This gives $\bk{K_2}$ a physical meaning as (part of) the one-reggeon
amplitude.

The second example is pion-nucleon elastic scattering in the large-$N_c$
limit. The tree diagrams are shown in Fig.~(d),(e), with pion-quark interaction
given by a Hamiltonian of the form 
$H_I\sim q^\dagger\v\Gamma\!\.\!\v\pi q$,
where $\v\Gamma$ is a matrix that contains the isospin and spin
information. The effective coupling at each vertex is proportional to
$\bk{H_I}/\sqrt{N_c}$, where $\bk{\cdots}$ represents the nucleon
matrix element and $1/\sqrt{N_c}$ is the normalization factor 
needed for each
external pion. Since a color-singlet nucleon contains $N_c$ quarks,
$\bk{H_I}\sim N_c$, so the effective coupling at each vertex is of order
$\sqrt{N_c}$, making each tree amplitude $\sim N_c$. However, the
term proportional to $N_c$ is canceled in the sum of the two
diagrams, leaving a total
amplitude $\bk{U_2}$ 
of order unity in the large-$N_c$ limit. This can be understood \cite{LL}
directly from \eq{k12} firstly because $\bk{K_1}=0$ on account of 
energy-momentum conservation: an on-shell nucleon cannot absorb or emit a
massive pion and remains on-shell. Thus $\bk{U_2}=\h\bk{K_2}$. 
Since the commutator of two one-body operators is again a one-body
operator, hence $\bk{[H_1,H_2]}\sim\bk{q^\dagger[\v\Gamma_1\!\.\!\v\pi_1,
\v\Gamma_2\!\.\!\v\pi_2]q}\sim N_c$.
Taking into account the normalization
factor $1/\sqrt{N_c}^2$ for two pions, we conclude that 
$\bk{U_2}=\h\bk{K_2}\sim N_c/N_c=1$, as needed. Like the situation in
Example 1, again there is no need to cancel the $N_c$ term explicitly,
because the commutator structure of $K_2$ already provides for it.
Physically, the one-body nature of $K_2$ tells us that the two pions
must interact with the same quark in the nucleon.

With these two examples, it is clear that the commutator structure
of $K_2$ is the key to the simplification. 
The question is whether $K_n$ still possesses
 such commutator structures for higher $n$. The answer is `yes', though
the detailed structure is increasingly more complicated. 
Let us define anti-hermitean
operators $C_n$ to have the simple nested commutator structure, 
\be
C_n&=&(-i)^n\int_{{\cal R}_n}d^nt\ [H_1,[H_2,[\cdots,[H_{n-1},H_n]
\cdots]]],\labels{cn}\ee
where $H_i\equiv H_I(t_i)$. Then it can be shown \cite{JMP} that
\be
U_n=\sum_{(m)}\prod_{j=1}^k{1\over\sum_{i=j}^k m_i}C_{m_1}C_{m_2}\cdots C_{m_k},
\labels{xic}\ee
where the sum is taken
over all partitions $(m)=(m_1m_2\cdots m_k)$
of the number $n$. In particular,
$U_1=C_1=K_1$,
$U_2=\(C_1^2+C_2\)/2=\(K_1^2+K_2\)/2$,
$U_3=\(C_1^3+C_1C_2\)/6+\(C_2C_1+C_3\)/3
=K_1^3/6+\(K_1K_2+K_2K_1\)/4+K_3/3$.
From these relation, or more generally by comparing \eq{unk} with 
\eq{xic}, one obtains the commutator structure of $K_n$ to be
$K_n=C_n+R_n$, with
$R_1=R_2=0$, 
$R_3=[C_2,C_1]/4$, and
$R_4=[C_3,C_1]/3$.
In general, $R_n$ is given by commutators of $C_m$, with increasingly
complicated coefficients as $n$ increases.

We can use these relations between $U_n$ and $C_m$ to calculate quark-quark
and electron-electron scatterings to higher orders, avoiding delicate
cancelations, and obtaining directly the energy dependence necessary for
reggeon structures in QCD, {\it viz.,} an amplitude proportional
to $g^{2k}(g^2\ln s)^{n-k}G_k$ when $k$ color-octet objects are being
exchange in the $t$-channel \cite{FHL,FDKL}.  
They can also be used to obtain the
correct amplitude \cite{LARGEN} $\sim N_c^{1-n/2}$
for $pi$-nucleon inelastic scattering with $n-1$ pions
in the final state, 
 without 
having to deal with delicate
cancelation of $n-1$ powers of $N_c$
 encountered by summing Feynman tree diagrams \cite{LL}. 
As mentioned at the beginnig, abelian and non-abelian
eikonal approximations are special cases of these formulas. 
Infrared structure of QED can 
be obtained from these formulas, with $K_1$
containing the Bloch-Nordsieck result, $K_2$ giving
rise to the Coulomb phase,
and $K_n=0$ for $n\ge 3$ \cite{IR,DKML}.
The Landau-Pomeranchuk-Migdal effect \cite{LPM}
can also be cast into this form \cite{DL}
with $K_n=0$ for $n\ge 3$. Other applications may require
knowing the the diagrammatic rules for calculating $K_n$.
For QED such rules are known \cite{DKML}.
For tree diagrams in QCD involving an energetic particle such rules
are also known \cite{FHL}, but the general case for QCD is still to be
worked out.

So far we have emphasized situations when $U$ is the time-evolution 
operator with a known Hamiltonian. If the dynamics is unknown, we can
still use $K_n$ in \eq{unk}, or $C_n$ in \eq{xic}, to parametrize the
{\it unitary} dynamics. As long as $K_n$ is anti-hermitean, 
it is clear that
$U$ is unitary. As long as $C_n$ is 
anti-hermitean, it can also be shown \cite{LF} that
 $U$ is unitary, though the proof is much more
involved. A Wolfenstein-like
parametrization of the CKM matrix to all orders can be deduced
from these parametrizations. They can also be used
to study the unitary matrix describing the overlap of the unperturbed
and the perturb energy eigenfunctions \cite{LF}. 

In conclusion, we have given a number of examples to show
that the unitary perturbation theory, in which
$K=\ln U$ is expanded perturbatively, is in many ways better than the
ordinary perturbation theory, where $U$ is directly expanded. 

This research is supported in part by the Natural Sciences and
Engineering Research Council of Canada, and the Fonds pour la
formation de Chercheurs et l'Aide \`a la Recherche of Qu\'ebec.
I am indebted to  Saad Ali, Hong-Mo Chan,
Marc Grisaru, Greg Mahlon, Sheung-Tsun Tsou,
Wu-ki Tung, and Kenneth Young for stimulating discussions.

\begin{figure}
\vskip 1 cm
\centerline{\epsfxsize 3 truein \epsfbox {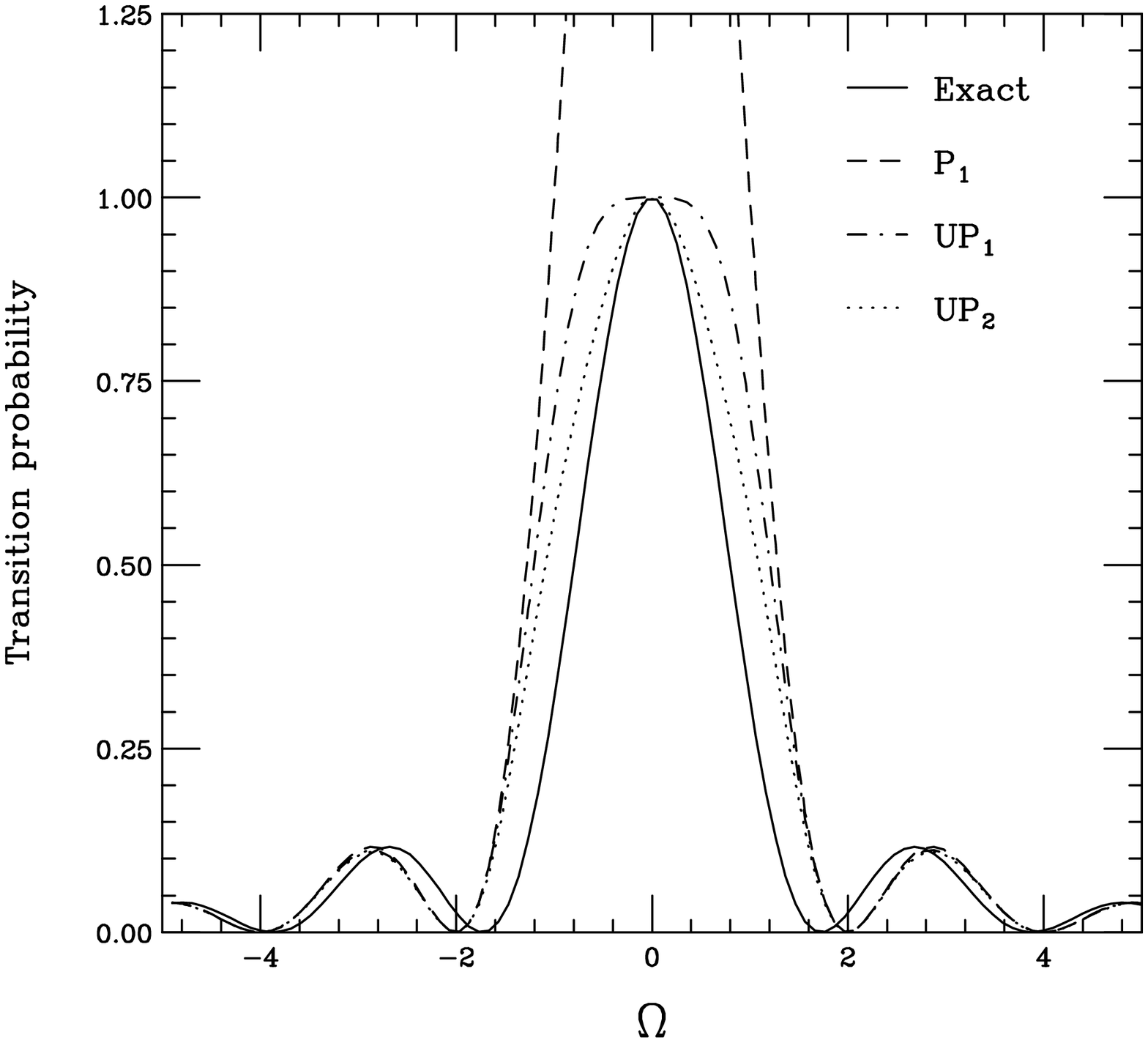}}
\nobreak
\vskip 0 cm\nobreak
\vskip .1 cm
\caption{Unitary perturbation theory vs ordinary perturbation theory
in a two-level example.}
\end{figure}

\begin{figure}
\vskip -2 cm
\centerline{\epsfxsize 3 truein \epsfbox {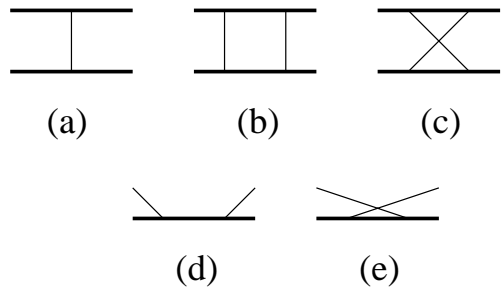}}
\nobreak
\vskip -.7 cm\nobreak
\vskip .5 cm
\caption{(a,b,c), high-energy quark-quark scattering diagrams to one loop
order.
(d,e), meson-nucleon scattering at large $N_c$ in the tree 
approximation .}
\end{figure}

\end{document}